\documentclass[twocolappendix]{emulateapj}

\usepackage{amsmath,amssymb,amstext}
\usepackage{mathtools}
\usepackage{gensymb}
\usepackage[usenames,dvipsnames]{color}
\usepackage{tabularx}
\newcommand\T{\rule{0pt}{2.6ex}}       % Top strut
\newcommand\B{\rule[-1.2ex]{0pt}{0pt}}
\newcolumntype{Y}{>{\centering\arraybackslash}X}
\usepackage{threeparttable}
\usepackage{ulem}
\usepackage[colorlinks,urlcolor=blue,citecolor=blue,linkcolor=blue,pdfusetitle]{hyperref}

\newcommand{\nickel}{\ensuremath{^{56}{\rm Ni}}}
\newcommand{\msun}{\ensuremath{M_{\odot}}}
\newcommand{\per}[1]{#1\ensuremath{^{-1}}}

\usepackage{comment}
\usepackage{todonotes}
\shorttitle{A single engine for GRB-SNe} 
\shortauthors{Barnes et al.}

\begin{document}

\title{A GRB and Broad-lined Type Ic Supernova from a Single Central Engine}

\author{Jennifer Barnes$^1$, Paul C. Duffell$^2$, Yuqian Liu$^3$, Maryam Modjaz$^3$, Federica B. Bianco$^{4,3}$, Daniel Kasen$^{1,2,5}$, Andrew I. MacFadyen$^4$}
\affil{$^1$ Department of Physics, University of California, Berkeley, CA 94720}
\affil{$^2$ Department of Astronomy and Theoretical Astrophysics Center, University of California, Berkeley, CA 94720}
\affil{$^3$ Center for Cosmology and Particle Physics, New York University, New York, NY 10003}
\affil{$^4$ Center for Urban Science and Progress, New York University, Brooklyn, NY 11201}
\affil{$^4$ Nuclear Science Division, Lawrence Berkeley National Laboratory, 1 Cyclotron Road, Berkeley, CA 94720}
\email{jlbarnes@berkeley.edu}

\begin{abstract}
Unusually high velocities ($\gtrsim 0.1c$) and correspondingly high kinetic energies have been observed in a subset of Type Ic supernovae (so-called ``broad-lined Ic'' supernovae; SNe Ic-BL), prompting a search for a central engine model capable of generating such energetic explosions. 
A clue to the explosion mechanism may lie in the fact that all supernovae that accompany long-duration gamma-ray bursts belong to the SN Ic-BL class.
Using a combination of two-dimensional relativistic hydrodynamics and radiation transport calculations, we demonstrate that the central engine responsible for long gamma-ray bursts can also trigger a SN Ic-BL. 
We find that a reasonable gamma-ray burst engine injected into a stripped Wolf-Rayet progenitor produces a relativistic jet with energy ${\sim} 10^{51}$ ergs, as well as a SN whose synthetic light curves and spectra are fully consistent with observed SNe Ic-BL during the photospheric phase. 
As a result of the jet's asymmetric energy injection, the SN spectra and light curves depend on viewing angle. The impact of viewing angle on the spectrum is particularly pronounced at early times, while the viewing angle dependence for the light curves (${\sim} 10$ \% variation in bolometric luminosity) persists throughout the photospheric phase.

\end{abstract}

\keywords{hydrodynamics --- shock waves --- instabilities --- supernovae: general --- ISM: jets and outflows }

\section{Introduction} \label{sec:intro}

The discovery of the supernova (SN) 1998bw \citep{Galama98_bwDisc} in apparent conjunction with the long-duration gamma ray burst (GRB) 980425 suggested a connection between long GRBs and SNe. The relationship was more firmly established with the detection of SN 2003dh rising out of the afterglow of GRB 030329 \citep{Stanek.ea_20032003dh.data,Hjorth.ea_2003_2003dh.disc.nature}, and has since been cemented by additional observations of SNe coincident with long GRBs. 
Notably, every SN linked to a GRB has been classified as a broad-lined SN Ic  \citep[Ic-BL; e.g.,][]{WoosleyBloom06_snegrbRev,Modjaz11_stellarFor,Cano.ea_2017_SN.GRB.Review}, a category of SNe whose
broad spectral features indicate high photospheric velocities \citep[$>20,000$ km \per{s};][]{Modjaz_etal_2016_GrbSne}.
The kinetic energies ascribed to these SNe are also high (${\sim} 10^{52}$ ergs), though they depend on the model used to compute 
the SN explosion parameters \citep[e.g.,][]{Mazzali.ea_2017_TypeIc.modeling}.

The search for an engine powerful enough to explain SNe Ic-BL energetics is one facet of an investigation into the explosion mechanism for core-collapse supernovae (CCSNe) more generally. 
In the traditional core-collapse theory, neutrino irradiation from the newborn neutron star revives the stalled shock launched at core bounce, and explodes the star \citep{ColgateWhite_1966_NeutrinoMech,BetheWilson_1985_NeutrinoMech}. 
SNe Ic-BL complicate this picture, since even in the most optimistic simulations, the power supplied by neutrinos is insufficient to explain the high kinetic energies of these extreme SNe.  

The link between SNe Ic-BL and long GRBs identified jets as a potential alternative source of explosive kinetic energy.
However the association of GRBs with SNe only of the Ic-BL class raises the question of whether jets, if they do indeed serve as engines, operate only in unusually energetic SNe, or are far more common events that could power a broad range of supernova explosions depending on the engine parameters and the nature of the progenitor star \citep{Sobacchi17b_grbSNIb,Sobacchi17a_SNjetEngine,Piran17_reljetsCCSNe}. 
Understanding the role of jets in extremely energetic explosions is an important step toward understanding the diversity (or lack of diversity) of CCSNe explosion mechanisms.

While all GRB-SNe are Ic-BLs, SNe Ic-BL have also been observed without coincident GRBs. Some SNe Ic-BL must accompany GRBs that point away from the line of sight, and thus go undetected, but it is not clear what fraction of SNe Ic-BL without observed GRBs are explained by orientation effects. In particular, radio follow-up of some GRB-absent SNe Ic-BL found no evidence of highly relativistic material  \citep{Berger_2002ap,SoderbergEtAl2006_sneIbc_offaxisgrb, CorsiEtAl2016_RadioQuietBlics}, suggesting that these SNe do not harbor off-axis GRBs.

A statistical analysis of SNe Ic-BL spectra \citep{Modjaz_etal_2016_GrbSne} revealed that GRB-SNe have systematically broader features than those SNe Ic-BL for which no GRB was detected.
There are three likely explanations for this trend.

The first is that SNe Ic-BL occur with \textit{and} without coincident GRBs, but the GRB jet, when present, increases the kinetic energy of the ejecta, resulting in higher photospheric velocities and broader spectral features. If this interpretation is correct, many SNe Ic-BL without an observed coincident GRB truly are solo explosions (though some will host GRBs that go unobserved because they point away from the line of sight). In this case,
the lower energies and narrower features of GRB-absent SNe Ic-BL observed by \citet{Modjaz_etal_2016_GrbSne} can be attributed to the fact that many of these SNe lack an energy-boosting GRB.

The second explanation is that the correlation between GRBs and high SN velocities is due to an as-yet undetermined third factor that produces both highly kinetic supernovae and relativistic jets. For example, it has been argued that rapidly rotating progenitors \citep[e.g.][]{Wheeler.ea_2002_jet.rotation.sn,
Uzdensky.MacFadyen_2006_Magnetic.Towers,  Burrows.ea_2007_sn.rotation} may allow the formation of both GRBs and high-velocity SNe.

A third possibility is that the jet induces an asymmetric explosion. In this case, the SN spectrum and photometry depend on viewing angle, and some of the distinction between the spectra of SNe Ic-BL with and without observed GRBs is due to line-of-sight effects. 
Specifically, the anisotropic energy injection from the jet engine could accelerate ejecta on or near the jet axis to higher velocities than material located at lower lattitudes. An observer looking down the barrel of the jet would see broader lines---and would infer a faster photosphere and a higher kinetic energy---than an observer viewing the system from an off-axis vantage \citep[e.g.,][]{Tanaka2007_MultiDRT}. 

Aspherical explosions have long been invoked to explain some puzzling features of SNe Ic-BL photometry. \citet{Hoflich_1999_asphBLIc} argued that a supernova with an oblate ejecta, if observed from a near-polar viewing angle, would appear to be more luminous than it truly was.
They suggested that SN 1998bw, the original Ic-BL, did not have the unusually massive ejecta and high 
\nickel\ content and kinetic energy suggested by 1D models, but in fact fell within the range of normal Type Ic SNe. The authors speculated that an oblate ejecta could also produce broader spectral lines in the polar direction than the equatorial direction, but did not carry out radiation transport calculations to verify the theory.
Asymmetry was also studied by
\citet{Wollaeger.ea_2017_blic.asym.2002ap.theory},
who explored whether a 
``unipolar'' asymmetry in the distribution of \nickel\ within the SN ejecta could reproduce the photometry of the SN Ic-BL 2002ap. 

\citet{Nakamura_01_98bwLCspec} and, more recently, \citet{Dessart2017_asymBLics} found that one-dimensional models are insufficient to explain the typical time evolution of Ic-BL photometry. The photospheric-phase light curves seem to require lower ejecta and \nickel\ masses and higher kinetic energies, while models of the late-time luminosity favor more a massive ejecta, lower amounts of \nickel, and lower kinetic energies. The schemes contrived to satisfy both the early- and late-time photometric constraints (e.g., the suggestion by \citet{Maeda03ApJ_twoCompNi_blic} that the distribution of \nickel\ in the Ic-BL ejecta is bimodal, with a high-velocity component powering the rapid rise, and a low-velocity, high-density component sustaining the light curve at late times) are implausible for a spherically symmetric ejecta, but, as both authors suggest, may be accommodated by aspherical explosions.

The question of asymmetry in SNe Ic-BL is especially vital given the on-going debate as to the central engine of these explosions. 
\citet{ThompsonChangQuataert_2004_Magnetars} outlined how energy extracted from short-lived magnetars produced by a supernova could modify the supernova shock dynamics, resulting in a hyper-energetic explosion.
The apparent clustering of the kinetic energies of observed SNe Ic-BL near the maximum rotational energy of a neutron star (${\sim}10^{52}$ ergs) led \citet{Mazzali2014_Magnetars} to suggest that magnetars are the kinetic energy sources of SNe Ic-BL, though there is open debate about how well a magnetar hypothesis explains the photometric properties of individual SNe Ic-BL \citep[see, e.g.,][]{WangEtAl2017_magnetarBlics,CanoEtal17_SN2016jca}.

\citet{MacFadyenWoosley99_collapsar} propose an alternate scenario in which the SN ejecta is blown off a disk accreting onto the black hole produced by core collapse. It has also been argued that the SN is driven by a jet engine, which also launches a GRB \citep{Maeda_nucleosynth_jetdriv, Sobacchi17a_SNjetEngine,Piran17_reljetsCCSNe}.

Central engine models, especially the magnetar model, are often evaluated based on their ability to produce explosions with kinetic energies close to the (presumed) canonical Ic-BL value of $10^{52}$ ergs. However, kinetic energies inferred from observations depend on parametrized 1D explosion models, which are not guaranteed to accurately map to explosions with significant asymmetry.  Fully relativistic hydrodynamical calculations of the explosion evolution, along with multi-dimensional radiation transport simulations of the resulting ejecta, can help resolve the question of Ic-BL energetics, and facilitate the development of more reliable tools for diagnosing supernova energies. This may lay the foundation for a more rigorous assessment of various engine models.

This paper will explore these questions in the context of a single, jet-driven explosion model. We will evaluate the effect of a GRB jet engine on a progenitor star, absent any other source of explosive energy, and demonstrate that such an engine, as it tunnels through the progenitor, can transfer sufficient energy to the surrounding stellar material to unbind it, naturally producing a supernova with a high kinetic energy. Additional energy from the engine escapes through the tunnel drilled in the star as an ultra-relativistic jet, and is observed as a GRB.

We perform a two-dimensional special relativistic hydrodynamic (SRHD) calculation, making it possible to predict asymmetries in the ejecta.  We find both the ejecta density profile and distribution of \nickel\ are aspherical. Two-dimensional radiation transport calculations allow us to track the effect of these asymmetries on the SN light curves and spectra. This is the first study to carry out an end-to-end (hydrodynamics and radiation) simulation of a jet-driven energetic SN in multiple dimensions.

The numerical tools used to simulate the hydrodynamics and radiation transport of the jet-SN system are outlined in \S~\ref{sec:numerics}. We define our engine and stellar progenitor models in \S~\ref{sec:models}. The resultant outflow is described in \S~\ref{sec:gasdynamics}.  The synthetic light curves and spectra of the SN, including viewing-angle dependence, are presented and \S~\ref{sec:photometry}.

\section{Numerical Methods} \label{sec:numerics}

We use a suite of advanced numerical tools to model the hydrodynamics and radiation of 
the jet-SN system, and to analyze the emergent SN spectra  and light curves. This suite allows us to study multidimensional supernova dynamics at a level of accuracy and efficiency ordinarily unavailable beyond 1D. 

\vspace{\baselineskip}
\noindent \textit{Hydrodynamics:} 
Hydrodynamical calculations are carried out using the \texttt{JET} code \citep{Duffell_2013_JET}, an efficient and accurate solver for the equations of relativistic fluid dynamics. \texttt{JET} employs a ``moving mesh" technique \citep{Duffell11_TESS_movmesh}, 
which makes it effectively Lagrangian in the radial dimension, while accurately evolving multidimensional flows, especially flows which move radially outward.  Among other advantages, the moving mesh makes it straightforward to evolve flows over large dynamic ranges.  

We use \texttt{JET} to inject energy and momentum into the progenitor star, following the method outlined in \citet{Duffell15_jetInject}, and assuming axisymmetry. This method, employed previously by \citet{Duffell15_jetInject} and \citet{Duffell.Quataert.MacFadyen_2015_narrow.jet.wide.engine} results in a robust evolution of the hydrodynamics, as the source terms are smooth functions which can be well-resolved, and the method does not require any special boundary conditions.
The engine injects into the core (on length scales ${\sim} 10^{8}$ cm) highly relativistic material with an energy-to-mass ratio of $10^3$. 

The subsequent hydrodynamical evolution of the fluid is followed for the full duration of the jet engine and proceeds until the flow becomes homologous (i.e., the gas coasts on ballistic trajectories with $\mathbf{v}(\mathbf{r},t) = \mathbf{r}/t$).

In actuality, the flow becomes homologous in stages, with the ultra-relativistic GRB jet reaching homology later than lower-velocity material.
However, the SN ejecta is only mildly relativistic ($v \lesssim 0.2c$), and radiation transport calculations of the SN will be unaffected by any late-time non-homology in the GRB jet.
We conclude the hydrodynamic phase when material with $v \lesssim 0.9c$ has reached homology, at which point the material comprising the supernova ejecta can be safely assumed to be homologous. 

This generally occurs at $t \sim$ few hours in physical time or, equivalently, an expansion to ${\sim} 10^3$ times the initial radius of the progenitor star.
Throughout the hydrodynamic phase, the material is relativistically hot and extremely optically thick, so the dynamical effects of radiation are fully contained in the choice of adiabatic index. 

The hydrodynamic calculation is used to determine the mass density profile of the ejecta and to approximate the synthesis of radioactive $^{56}$Ni. 
Our model does not include a detailed nuclear reaction network, so we estimate the production of $^{56}$Ni 
with a simple temperature condition. Any zone in which the temperature exceeds $5 \times 10^9$ K is assumed to burn to pure \nickel. The \nickel\ mass fraction is advected along with the flow, allowing synthesized \nickel\ to spread through the ejecta.
The synthesis and subsequent decay of \nickel\ release energy, but the energy is negligible compared to the thermal and kinetic energies of the fluid during the hydrodynamical phase. 
We therefore assume that nuclear energy release is not important for the hydrodynamical evolution of the system. (The decay of \nickel\ is the source of luminosity for the supernova, and radioactive energy \textit{is} included in the radiation transport calculation.)

\vspace{\baselineskip}
\noindent \textit{Radiation Transport:}  We perform radiation transport with \texttt{Sedona} \citep{Kasen_MC}, a time-dependent multi-wavelength, multi-dimensional Monte Carlo radiation transport code that has been used extensively to model radioactively-powered transients, in particular SNe.
Radiation transport is performed on a two-dimensional axisymmetric grid constructed from the low-velocity region of the ejecta structure output by \texttt{JET}. We take as the boundaries of the grid $v_s = |v_z| = 0.2c$. We found that extending the grid to higher velocities had no appreciable effect on the results of the radiation transport. 

\texttt{Sedona} evolves the density and temperature of the ejecta, and accounts for radioactive heating by the decay of $^{56}$Ni and $^{56}$Co, and cooling by expansion. The ionization and excitation states in the ejecta are determined assuming local thermodynamic equilibrium (LTE), and detailed wavelength-dependent opacities are calculated from the atomic line lists of \citet{Kurucz_CD23,Kurucz_Bell_1995}. Line opacity is assumed to be completely absorptive. \texttt{Sedona} synthesizes a full spectral time series, from which we construct the emergent light curves and spectra of the supernova.

\vspace{\baselineskip}
\noindent \textit{Spectral Analysis:}
We analyze our synthetic spectra using state-of-the-art tools developed for the most statistically rigorous study of observed GRB-SNe to date.
We compare our synthetic spectra to the average spectra of SNe Ic-BL computed by \citet{Modjaz_etal_2016_GrbSne} at different phases, allowing us to test the agreement of the model with the full diversity of SNe Ic-BL, rather than simply comparing to individual objects.
The velocities implied by the model spectra are determined with template-fitting methods \citep[see][Appendix A]{Modjaz_etal_2016_GrbSne} used
to characterize the spectra of observed SNe Ic-BL. 
We also perform spectral analysis with the Supernova Identification code \citep[SNID;][]{Blondin.Tonry_2007_SNID.paper}, using a full library of SNID spectral templates supplied in part by \citet{Liu.Modjaz_14_sn.templates,Modjaz_etal_2016_GrbSne,Liu.Modjaz.ea_2016_se.sn.spectra,Liu.Modjaz.Bianco_2016arX_Hpoor.SlSne}.\footnote{Release via https://github.com/nyusngroup/SESNtemple}
SNID allows us to determine statistically the spectroscopic category that best describes our model, and to find observed supernovae that best match the synthetic model spectra. 
\section{Progenitor and Engine Models}
\label{sec:models}

\noindent \textit{Stellar Progenitor Model:}
SNe Ibc are generally assumed to be the explosions of stripped-envelope Wolf-Rayet (WR) stars, or of stars in binary systems whose masses are lower than expected for WR stars. (However, whether the WR progenitor scenario can explain SNe Ic-BL is a topic of active research; see \citet{Dessart2017_asymBLics} for a full discussion.)
In this work, we use an analytic progenitor model that reasonably approximates the major features of a WR star. Future work will use detailed WR progenitor models evolved with the stellar evolution software \texttt{MESA} \citep{Paxton15_MESA}.

We assume that the innermost regions of the progenitor collapse to a compact object. We therefore remove from the computational grid material interior to $r_{\rm cav} = 1.5 \times 10^{-3} R_{\odot} \approx 1000$ km. The mass of the excised material is $\sim 1.4 \msun$. For numerical tractability, the density in the cavity region is set to $10^{-3}$ times the density at the cavity boundary.

The material at $r > r_{\rm cav}$ is assumed to be unaffected by the collapse, and thus to maintain the pre-collapse stellar density profile. The density of this material, $\rho_{\rm init}$, is a function of radius only,
\begin{align}
\rho_{\rm init}(r) = \frac{0.0615 M_0}{R_0^3} (R_0/r)^{2.65}(1-r/R_0)^{3.5}.
\end{align}
Above, $R_0 = 1.6 R_\sun$ is the radius to which the stellar atmosphere extends, and $M_0 = 2.5 \msun$ sets the mass of the material outside the cavity. Including the mass of the central remnant (${\sim} 1.4 \msun$), our model suggests a stellar mass at collapse of $\lesssim 4 \msun$.

\begin{figure}
\includegraphics[width=3.5 in]{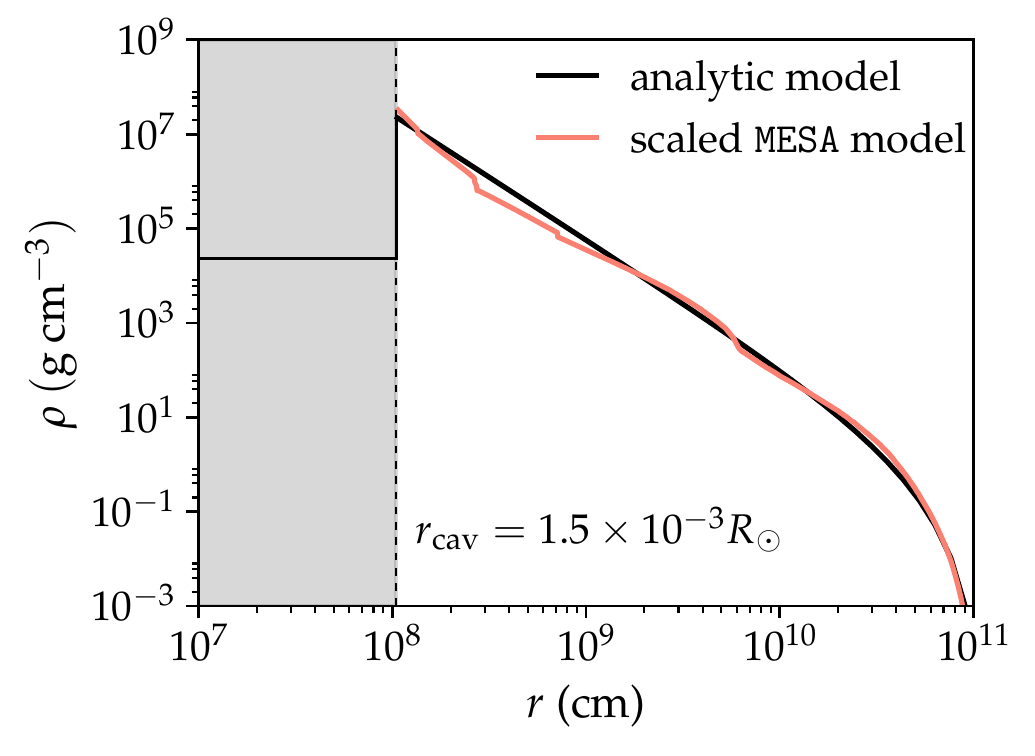}
\caption{The analytic density profile of the progenitor star after core collapse (solid black curve). We assume the collapse of the core leaves behind a cavity of radial extent $r_{\rm cav}$. The density in this region is set to a low, constant value. For comparison, we also plot the scaled density profile, for $r > r_{\rm cav}$, of a stripped-envelope star evolved with \texttt{MESA} (red curve). The analytic model captures the major qualitative features of the \texttt{MESA} result.} 
\label{fig:progstar}
\end{figure}

The analytic progenitor model is shown in Figure \ref{fig:progstar}. For comparison, we also show the density profile, for $r > r_{\rm cav}$, of a WR star evolved with \texttt{MESA}, which we have scaled so that the mass exterior to $r_{\rm cav}$ totals 2.5 \msun, as in our model. 
While actual WR stars will not follow these simple scaling laws, the qualitative features of the density profile outside the Fe core are unlikely to change dramatically with stellar mass, and the core collapse, in our calculation, is imposed \textit{a priori}, rather than generated self-consistently, and so is insensitive to assumptions about the core density structure. 
One disadvantage of this approach is that it does not rigorously determine the density in the innermost regions where most of the \nickel\ is synthesized. This introduces an additional uncertainty into our estimate of \nickel\ production, which is discussed further in \S~\ref{sec:gasdynamics} .

Table \ref{tab:sum} presents the composition, as mass fractions, of the progenitor star.
Oxygen dominates, as might be expected for a stripped-envelope Carbon-Oxygen star. 
The ejecta composition, modulo \nickel, is taken to be spatially uniform;
in zones containing $^{56}$Ni, the composition is scaled to accommodate the \nickel\ content, while preserving the relative mass fractions of all non-radioactive species.
Future work will use more realistic compositions that better reflect compositional inhomogeneities found in detailed studies of evolved stars and CCSN progenitors \citep[e.g.,][]{Arnett.Meakin_2011_CCSNe.progen, Couch.ea_2015_CCSNe.progen.Ev.3d, Chatzopoulos.Couch.ea_2016_Conv.Rot.SNprog.2d}.

\begin{table*}[ht]
\caption{Model Summary}\label{tab:sum}
\begin{threeparttable}
\centering
\begin{tabularx}{\textwidth}{*{12}{Y}}
\hline \hline
\multicolumn{6}{c}{\textit{Progenitor Parameters}} & \multicolumn{6}{c}{\textit{Engine Parameters}} \T\B \\
\hline 
\multicolumn{2}{c}{$M_\mathrm{tot}$\tnote{a}} & \multicolumn{2}{c}{$M_{\rm CR}$} & \multicolumn{2}{c}{$M(r > r_{\rm cav})$\tnote{b}} & \multicolumn{2}{c}{$E_{\rm eng}$} & \multicolumn{2}{c}{$t_{\rm eng}$\tnote{c}} & \multicolumn{2}{c}{$\theta_{\rm eng}$\tnote{d}} \T\B \\
\multicolumn{2}{c}{$3.9 \msun$} & \multicolumn{2}{c}{$1.4 \msun$} & \multicolumn{2}{c}{$2.5 \msun$} & \multicolumn{2}{c}{$1.8 \times 10^{52}$ erg} & \multicolumn{2}{c}{$1.1$ s} & \multicolumn{2}{c}{$11.5 \degree $} \\
\vspace{.3 mm} \\
\multicolumn{12}{c}{\textit{Progenitor Composition}} \B \\
\hline
He & C & N & O & Ne & Mg & Si & S & Ar & Ca & Ti & Fe \T \\ 
6.79e-3 & 2.27e-2 & 2.91e-5 & 9.05e-1 & 1.37e-2 & 8.46e-3 & 2.69e-2 & 1.04e-2 & 1.60e-3 & 6.63e-4 & 5.11e-7 & 3.50e-3 \T\B \\
\hline \hline
\end{tabularx}
\begin{tablenotes}
\item[a] The mass of the evolved progenitor star just prior to core collapse.
\item[b] The mass remaining after the core has been excised.
\item[c] This relatively short-duration engine generates a GRB of longer duration. See Section \ref{sec:gasdynamics} and Table \ref{tab:sngrb} for details.
\item[d] The eventual opening angle of the GRB jet is narrower than $\theta_\mathrm{eng}$ (see Table~\ref{tab:sngrb}), due to recollimation processes that act on the jet as it tunnels through the star.
\end{tablenotes}
\end{threeparttable}
\end{table*}

\vspace{\baselineskip}
\noindent \textit{Jet Engine Model:}
The GRB engine model is defined by the total energy injected, $E_{\rm eng}$; the engine half-opening angle, $\theta_{\rm eng}$; and the characteristic timescale of the engine, $t_{\rm eng}$. While $t_{\rm eng}$ is often assumed to be greater than or equal to the burst duration, we find in \S~\ref{sec:gasdynamics} that a short engine can produce a GRB of duration $\tau_\mathrm{GRB} > t_\mathrm{eng}$. 
Rather than cutting the jet luminosity off instantaneously, we allow it to decay
exponentially over the timescale $t_{\rm eng}$,
\begin{equation}
L_{\rm eng}(t) = \frac{E_{\rm eng}}{t_{\rm eng}} \times \exp[-t/t_{\rm eng}].
\end{equation}
The engine is symmetric about the equatorial plane, so $E_{\rm eng}$ is the sum of the energy injected along the positive and negative $z$-axis.

We focus on one engine model with $E_{\rm eng} = 1.8 \times 10^{52}$ ergs, $\theta_{\rm eng} = 11.5\degree$, and $t_{\rm eng} = 1.1$ s. 
(Note that the engine duration and opening angle are different from the GRB duration and the opening angle of the GRB jet, as discussed below.) 
These parameters were found to produce a SN and a GRB roughly consistent with observations; 
future work will more fully explore the dependence of the GRB and SN on the engine parameters.

\section{Gas Dynamics}
\label{sec:gasdynamics}

\begin{figure}
\includegraphics[width=3.5 in]{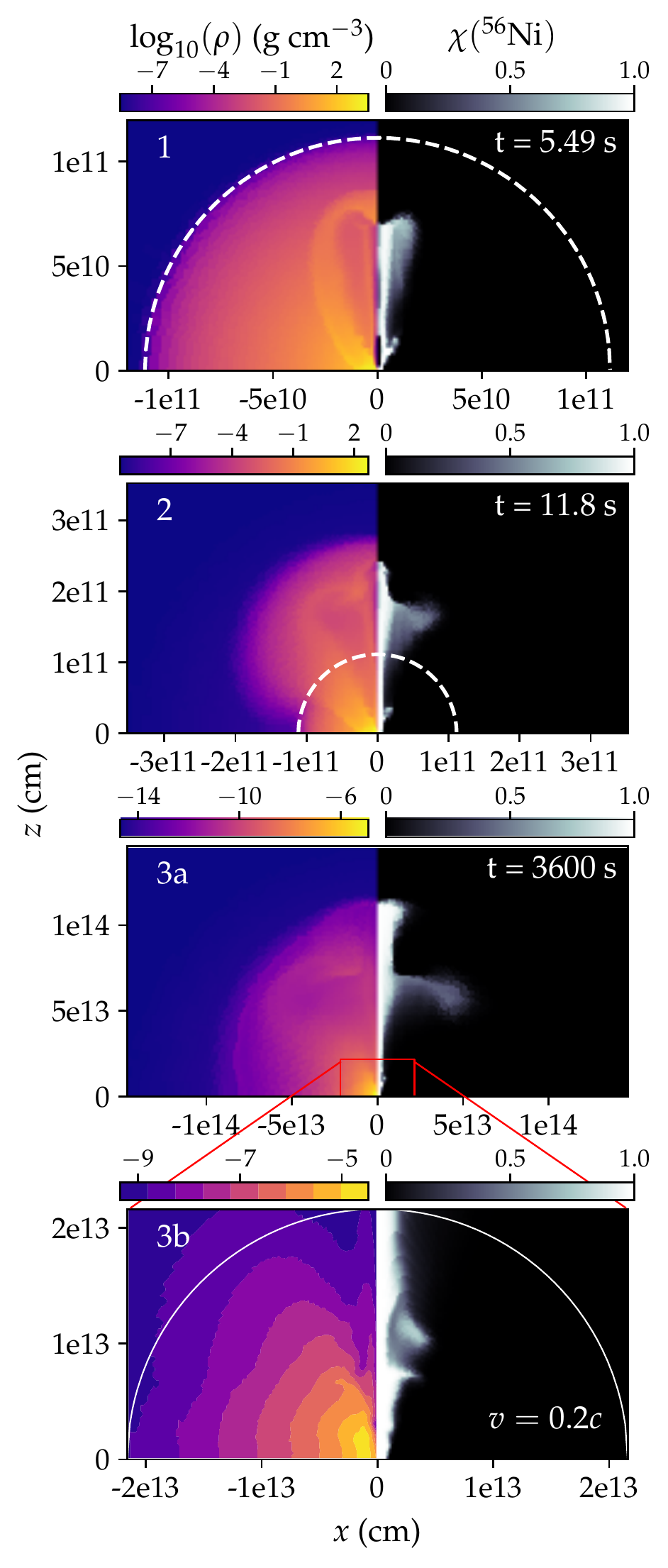}
\vspace{-10 pt}
\caption{The gas dynamics of the jet. The left (right) side of each panel shows mass density (\nickel\ mass fraction). The pre-explosion progenitor radius is plotted in Panels 1 and 2 as a dashed white line. The engine burrows through the progenitor (Panel 1), disrupts it, and eventually breaks out (Panel 2). Energy is transferred to off-axis material by lateral shocks.
In Panel 3a, the system is homologous. 
The most relativistic material has erupted as a GRB jet, but much of the matter reached lower velocities (a few $\times \: 0.1 c$), forming the supernova explosion. We perform our radiation transport calculations on material inside the red box of Panel 3a, which is defined by $v_s, \: |v_z| \leq 0.2c$. Panel 3b zooms in on this region, and shows density contours to emphasize the ejecta geometry. The white line in Panel 3b shows $v = 0.2c$.}
\label{fig:outflow}
\end{figure}

The jet propagation and \nickel\ production are illustrated in Figure \ref{fig:outflow}.
The engine injects energy into the center of the star, creating a hot, high-pressure region that tunnels along the $z$-axis toward the surface of the progenitor, which is marked by a dashed white line in the top two panels.

As the head of the jet burrows outward, the energy from the engine is redistributed throughout the star. Recollimation shocks confine the most relativistic material to a narrow region around the $z$-axis (see Panel 1), and shocks emanating from that hot, high-pressure region push against the cold stellar material off-axis, accelerating it to high, but non-relativistic, velocities (Panel 2). As the engine heats the stellar material to temperatures exceeding $5 \times 10^9$ K, \nickel\ is synthesized.  The highest temperatures occur near the center of the star.  The \nickel\ forged at small radii is then entrained by the relativistic flow propagating toward the pole and deposited in a narrow cone about the $z$-axis.

The relatively short duration of the engine ($1.1$ s) allows our model to produce a fairly high amount of \nickel\ (0.24 \msun). The energy scale of the engine is constrained by the observed kinetic energies of SNe Ic-BL. A shorter engine injects this energy in a concentrated burst, and drives up temperatures deep in the interior of the star before these inner layers can react to the energy injection and expand to lower densities. In our model, the densities in the zones that satisfy $T_{\rm max} \geq 5 \times 10^9$ K are high enough that a non-negligible amount of \nickel\ is synthesized, though the exact amount will be sensitive to the densities at very small radii, and will depend on the progenitor structure and the physics of the core collapse.
Models with longer-lived jets may have difficulty synthesizing similar quantities of \nickel; for example, the jets of \citet{ChenEtAl2017_BlicMagnetar} have a duration of ${\sim} 20$ s, but produce only $0.05\msun$ of \nickel.

\begin{figure}\includegraphics[width=3.5 in]{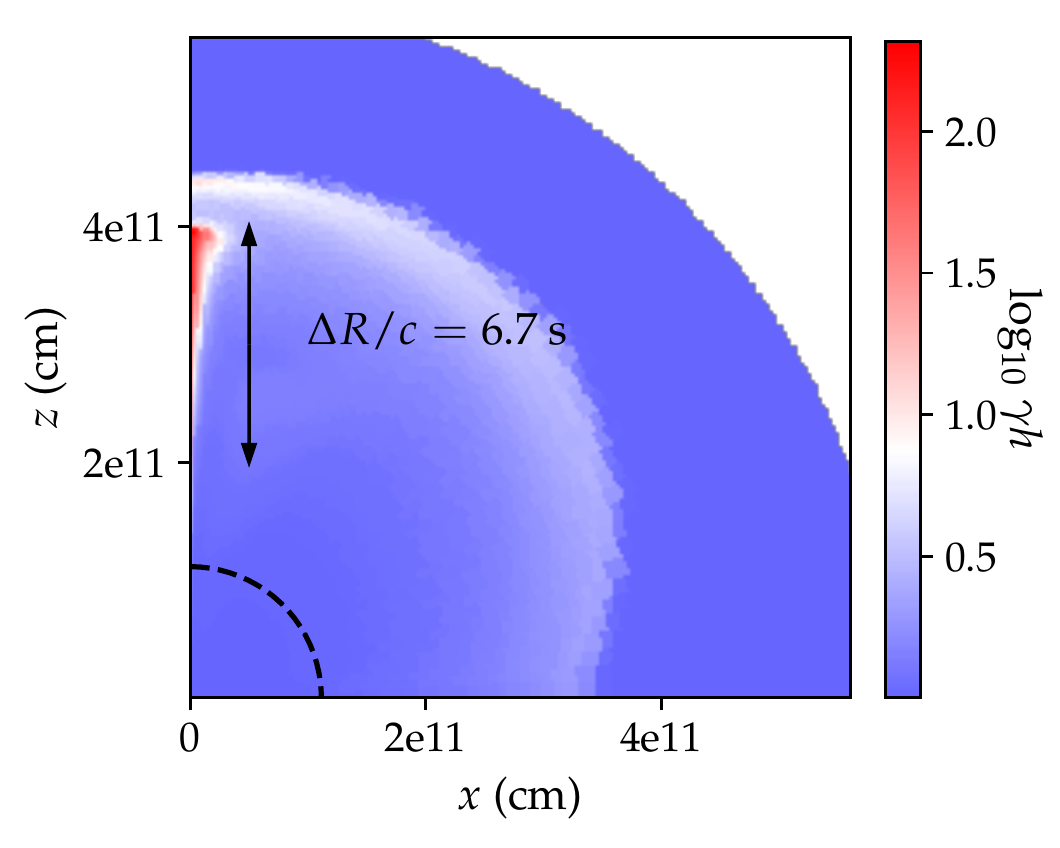}
\caption{The scaled terminal Lorentz factor, $\gamma h$ of the material, at $t=17.3$ s. At this time, the jet has broken out of the star (the initial stellar radius is plotted as a black dashed line) and the spatial scale of the stream with high $\gamma h$ is no longer evolving. The radial thickness of material with $\gamma h \gtrsim 10$ suggests a GRB duration of $\lesssim 7$ s, a factor of $>5$ greater than $t_{\rm eng}$. The flow does not attain Lorentz factors quite as large as shown above at late times (the peak late-time Lorentz factor is 72) because the most relativistic parts of the flow are eventually decelerated by internal collisions.}
\label{fig:gamH}
\end{figure}

The importance of high engine luminosities/short engine timescales for significant ($\gtrsim 0.1 \msun$) \nickel\ production was pointed out by \citet{Maeda_nucleosynth_jetdriv}. 
Here, we demonstrate that a short-duration engine can also produce an extended relativistic stream consistent with a long-duration GRB. Figure \ref{fig:gamH} shows
$\gamma h$, the local Lorentz factor multiplied by the specific enthalpy of the fluid. We have scaled $h$ by $c^{-2}$ to convert it to a dimensionless quantity. 
The product $\gamma h$ is fixed for an expanding fluid element. The dimensionless $h$ approaches 1 as the flow evolves, so the Lorentz factor $\gamma$ asymptotes to the initial value of $\gamma h$.
The timescale for prompt $\gamma$-ray emission is set by the width of the relativistic (high $\gamma h$) jet, which is fixed soon after the jet escapes the star.  If all the relativistic material were shocked instantly, an observer would detect a pulse lasting $\tau_{\rm GRB} = \Delta R/c$, where $\Delta R$ is the jet's radial thickness.

It is generally assumed that $\Delta R = c t_{\rm eng}$, based on the idea that the engine emits over its duration a stream of relativistic material moving at $c$. 
However, this picture does not take into account the process of the jet pushing its way through the stellar envelope, or the effects of the high-pressure cocoon supporting the jet.
At the point of breakout, the engine has created a stream of relativistic material extending from the engine to surface of the star, a distance $\Delta R = R_0 = 1.1 \times 10^{11}$ cm.
This stream flows through a low density tunnel drilled by the jet head as it propagates through the star. The tunnel sits inside a cocoon whose pressure, just before the jet head emerges from the star, is in rough equipartition with the jet's kinetic energy density.
After jet breakout, this pressure is converted into kinetic energy, increasing the size of the relativistic stream.
Once the cocoon pressure has dropped, the evolution of the stream ends. At this point, the relativistic stream has $\Delta R \approx 2 \times 10^{11}$ cm, corresponding to a GRB of duration $\tau_{\rm GRB} \approx 6.7$ s. Figure \ref{fig:gamH} shows the radial width of the relativistic material explicitly.

By the time homology is achieved (about an hour after the engine is initiated), the system has formed two distinct high-energy components.
The first is a highly relativistic GRB jet, which can be seen at the outer edge of the ejecta in Panels 2 and 3a of Figure~\ref{fig:outflow}. 
The GRB jet produced for our choice of engine and progenitor has a half-opening angle of $2.9\degree$ after breaking out of the star, an estimated duration of $\lesssim 7$ seconds, and a peak (average) Lorentz factor of 72 (40). \citep[Figure \ref{fig:gamH} shows $\gamma h$ values somewhat higher than this peak. This is because internal collisions eventually decelerate the most relativistic components of the flow; see][]{Duffell15_jetInject}. 
Recollimation shocks confine the highly-relativistic material in a narrow column; as a result, the GRB jet has an opening angle narrower than that of the engine.
The jet and counter-jet each have ${\sim} 2 \times 10^{51}$ ergs of kinetic energy, or roughly 10\% of the total engine energy.

The second component is a fairly isotropic SN explosion dominated by lower-velocity  ($v \lesssim 0.2c$) material, which is demarcated by a red box in Panel 3a and detailed in Panel 3b.
The SN ejecta has a mass of 1.9 \msun\ and a kinetic energy of $7.4 \times 10^{51}$ ergs. (Some of the engine energy is spent accelerating material that is not part of the GRB jet and has densities too low and velocities too high to contribute to the SN ejecta, so the energy in the GRB jets and the SN sum to less than $E_{\rm eng}$.) The properties of the GRB and SN are summarized in Table \ref{tab:sngrb}.

The SN ejecta is shown in detail in Panel 3b. There is a very narrow prolate component close to the $z$-axis,
surrounded by a torus with a roughly ellipsoidal cross-section.
Overall, the deviations from isotropy are minor.
The distribution of \nickel, which is concentrated along the jet axis, exhibits far more anisotropy, though we note that jet instabilities not captured in 2D might alter the distribution of \nickel\ further, possibly affecting the SN. Our model produces 0.24\msun\ of \nickel, though \nickel\ production depends sensitively on the parameters of the engine (see above), a theme that will be explored in future work.

\begin{table}[h]
\caption{SN and GRB Properties}\label{tab:sngrb}
\begin{threeparttable}
\begin{tabularx}{\columnwidth}{*{3}{Y}}
\hline \hline
\multicolumn{3}{c}{\textit{GRB Properties}} \T\B \\
\hline
$\tau_{\rm GRB}$\tnote{a} & $E_{\rm jet}$ & $\theta_{\rm jet, 1/2}$ \T\B \\
6.7 s & $2 \times 10^{51}$ ergs & $2.9$\degree \\
\vspace{.3 mm} \\
\multicolumn{3}{c}{\textit{SN Properties}} \T\B \\
\hline
$t_{\rm rise}$ & $L_{\rm peak}$ & $M(\nickel)$ \T\B \\
$17.5$ days & $6 \times 10^{42}$ erg s\per & $0.24 \msun$ \\
\hline \hline
\end{tabularx}
\begin{tablenotes}
\item[a] GRB duration is estimated as described in \S~\ref{sec:gasdynamics}
\end{tablenotes}
\end{threeparttable}
\end{table}

\section{Supernova Observables}
\label{sec:photometry}

We perform two-dimensional radiation transport calculations on the SN ejecta using the time-dependent multi-wavelength transport code \texttt{Sedona} \citep{Kasen_MC}. To assess the effects of viewing angle, we calculate light curves and spectra for seven evenly-spaced bins in $\mu = \cos\theta$, $\mu \in [-1,1]$. The spectra and light curves shown here are averages within the bins.

\subsection{Bolometric Light Curves} 
\label{subsec:bollum}

\begin{figure}\includegraphics[width=3.5 in]{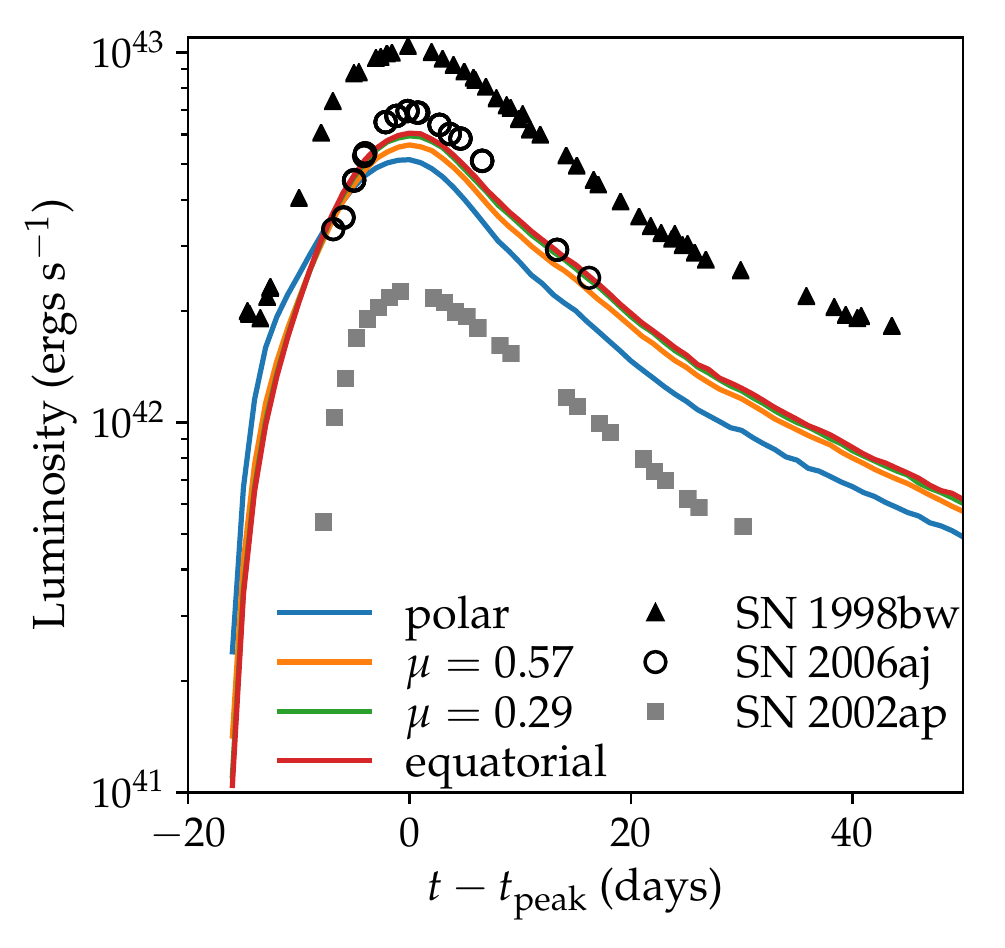}
\caption{Bolometric light curves of the model, for a range of viewing angles, compared to reconstructed bolometric light curves for three SNe Ic-BL from \citet{Prentice.ea_2016_SeSne.bolLCs}. The peak luminosities and widths of the model light curves are consistent with observations, though our model has a longer rise time than most observed SNe Ic-BL. The effect of viewing angle is modest, with observers near the pole seeing luminosities $\lesssim 10\%$ lower than viewers along the equator.}
\label{fig:bolLC}
\end{figure}

Figure \ref{fig:bolLC} shows the bolometric luminosity of our model SN for a range of viewing angles. (Due to symmetry about the equatorial plane, resolving the emission into seven bins in $\mu$ produces only four distinct light curves, and only four are presented.) For comparison, we also plot bolometric light curves for three SNe Ic-BL: 1998bw, which accompanied GRB 980425 \citep{Galama98_bwDisc}; 2006aj, which was coincident with GRB 060218B \citep{Mirabal06_06ajDisc,
Campana.ea_2006_2006aj.data, 
Modjaz.ea_2006_2006aj.data, Pian.ea_2006_2006aj.data,
Soderberg.ea_2006_2006aj.GRB060218.data,Sollerman.ea_2006_2006aj.data}; and 2002ap, which had no observed GRB and no indications of any relativistic outflow \citep{Berger_2002ap}. These pseudobolometric light curves were constructed by \citet{Prentice.ea_2016_SeSne.bolLCs}, based on data from references therein.

The synthetic light curves reach a peak bolometric luminosity of ${\sim} 6 \times 10^{42}$ ergs \per{s} at $t = 17.5$ days. This is slightly less than the average peak luminosity of SNe Ic-BL reported by \citet{Prentice.ea_2016_SeSne.bolLCs} ($\log L_{\rm peak} = 43 \pm 0.21$), but as shown in Figure \ref{fig:bolLC}, still falls within the range of observed Ic-BL light curves. The width and general shape of the light curves are also consistent with observations. However, our model has a longer rise time than many observed SNe Ic-BL.
For example, the SNe in Figure \ref{fig:bolLC} have $t_\mathrm{rise}$ $\sim 15$ days (1998bw), $\sim 9$ days (2006aj), and $\sim 12$ days (2002ap).
The values of $t_\mathrm{rise}$ for SNe Ic-BL and GRB-SNe calculated by \citet{Prentice.ea_2016_SeSne.bolLCs} are also generally lower than the model value of 17.5 days.
A different choice of engine and/or progenitor may resolve the rise time discrepancy. A full parameter space study is planned to investigate the effects of engine and progenitor properties on SN observables.

Surprisingly, orientation has  little effect on either rise time or luminosity; the brightest peak luminosity in our model is only ${\sim}10\%$ higher than the dimmest.
The anisotropic distribution of \nickel\ does not produce a strong viewing-angle dependence for the observed luminosity because the majority of the \nickel\ resides deep in the interior of the ejecta. (In our model ${\sim} 88\%$ of the \nickel\ mass is located at velocities less than $0.05c$.)
The small amount of \nickel\ at higher velocities imparts to the polar light curve a slightly higher luminosity at very early times ($t \lesssim 5$ days), but has a negligible effect on the light curve otherwise.

The energy radiated by the bulk of the \nickel\ must be reprocessed by the intervening material before it escapes the ejecta.
The mild dependence of luminosity on viewing angle exhibited by our model originates not in the \nickel\ distribution, but in the ejecta geometry.
As shown in Panel 3b of Figure~\ref{fig:outflow},
apart from a narrow region of lower-density material in the cone about the z-axis evacuated by the jet,
the ejecta is approximately spherical.
Small variations with $\mu = \cos \theta$ in the ejecta density, apparent in Panel 3b of Figure~\ref{fig:outflow}, produce the slight difference in bolometric luminosity seen in Figure~\ref{fig:bolLC}.
The trend of lower luminosity at higher latitudes was also found by \citet{Wollaeger.ea_2017_blic.asym.2002ap.theory}, though their model had much larger variation in $L_\mathrm{peak}$ due to a more pronounced asymmetry in the density and \nickel\ distributions.

\subsection{Spectra}
\label{sec:spectra}

\begin{figure}\includegraphics[width=3.5 in]{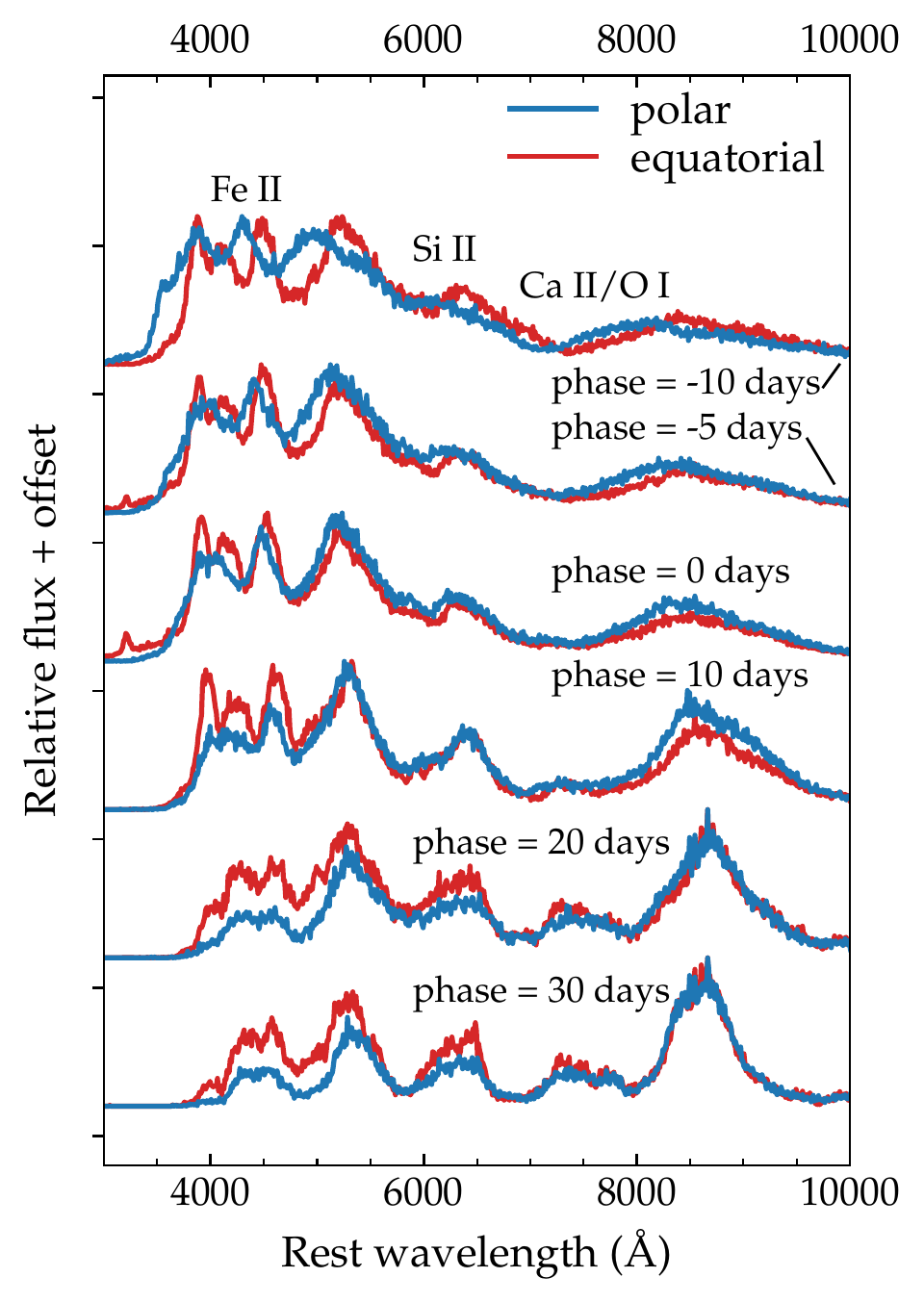}
\caption{The time-evolution of the spectra from our model supernova for polar (blue curves) and equatorial (red curves) viewing angles. Phases are relative to maximum light. We have annotated the topmost spectrum, indicating the ions that most contribute to the formation of spectral features in different regions of wavelength space.}
\label{fig:spec}
\end{figure}

The time evolution of the SN spectrum is shown in Figure \ref{fig:spec} for polar (blue curves) and equatorial (red curves) viewing angles. In all figures, the spectra are plotted as normalized flux per \AA, $F_\lambda$. 
We have indicated for the topmost spectrum in Figure~\ref{fig:spec} the ions that dominate the spectral formation in various regions of wavelength space.

Both viewing angles produce spectra whose flux at early times is concentrated at wavelengths 3500 \AA $\lesssim \lambda \lesssim$ 5500 \AA. As time progresses, the flux is redistributed redward and emission for $\lambda \lesssim 5000$ \AA\ becomes less pronounced.

For phases greater than 10 days, the spectra exhibit excess flux at  8500 \AA, associated with an emission line of CaII. 
This is due to the assumption of LTE in the radiation transport calculation, which is known to over-estimate the ionization fraction of Ca \citep{Kasen_IR}.

The effect of viewing angle is more obvious
for the spectra than for the light curves. The differences are particularly pronounced prior to bolometric peak. At phases $-$10 and $-$5 days, the spectra for the polar viewing angle show broader and bluer features than the equatorial spectra, especially at UV/blue wavelengths (3000 \AA $\; \leq \lambda \leq$ 5000 \AA). This reflects the asphericity of the photosphere, which forms at a slightly higher velocity along the poles than near the equator. 

For $t > t_{\rm rise}$, the flux at UV wavelengths is suppressed for the polar spectra relative to the equatorial spectra, which have more prominent features in the range 4000 \AA $\; \lesssim \lambda \lesssim$ 5000 \AA, and are bluer overall. These discrepancies are due to the presence of explosively-synthesized $^{56}$Ni/Co/Fe along the pole, which has numerous strong bound-bound transitions in the UV and increases the line opacity.

In order to identify the characteristic velocity scales of the model and compare them with those of observed SNe Ic-BL, we measured for the polar spectrum at maximum light the velocity of the Fe II triplet (i.e., Fe II $\lambda\lambda\lambda$ 4924, 5018, 5169 \AA) using the methods and code described in  \citet{Modjaz_etal_2016_GrbSne}, which were also used to analyze the largest dataset to date of observed SNe Ic-BL with and without GRBs. These template-fitting and broadening procedures properly account for the line-blending that occurs at the observed high velocities of SNe Ic-BL.

We find that the Fe feature complex is blueshifted by $v_\mathrm{blue} = -18,900 \substack{+4,200 \\ -4,100}$ km \per{s}.
We also calculate the ``convolution velocity,'' $v_\mathrm{conv}$, which represents the full width at half maximum (FWHM) of the Gaussian kernel that, when convolved with the narrow-lined SN Ic template, reproduces the width of the SN Ic-BL Fe II feature.
For our model, $v_\mathrm{conv} = 15,000 \substack{+7,500 \\ -14,800}$ km \per{s}. 
For comparison, the spectra of observed SNe Ic-BL (with and without GRBS) have an average $v_\mathrm{blue}$ of $-21,000 \pm 8,200$ km \per{s} (at maximum light), and a median $v_\mathrm{conv}$ of ${\sim} 8,000$ -- 9,000 km \per{s} (over all phases).

Thus, the model spectra are in good agreement with observed spectra,
though the error bars for the model's $v_\mathrm{conv}$ are quite large, reflecting some tension in the fitting algorithm.
The source of this tension is the shape of the Fe II feature. In normal SNe Ic, the triplet appears as a double trough with a pronounced bump in the center. In SNe Ic-BL, the central bump is often washed out by strong line blending \citep[see Appendix A of][]{Modjaz_etal_2016_GrbSne}. In our model, the ``W'' shape of the triplet is suppressed, but the total width of the blended feature is lower than for a typical SN Ic-BL. The fitting scheme favors higher $v_\mathrm{conv}$ to broaden away the central bump, but requires lower values to better capture the overall width, resulting in large uncertainties.
Overall, however, the best-fit values of $v_\mathrm{blue}$ and $v_\mathrm{conv}$ are consistent, within error bars, with those of observed SNe Ic-BL.

\begin{figure}\includegraphics[width=3.5in]{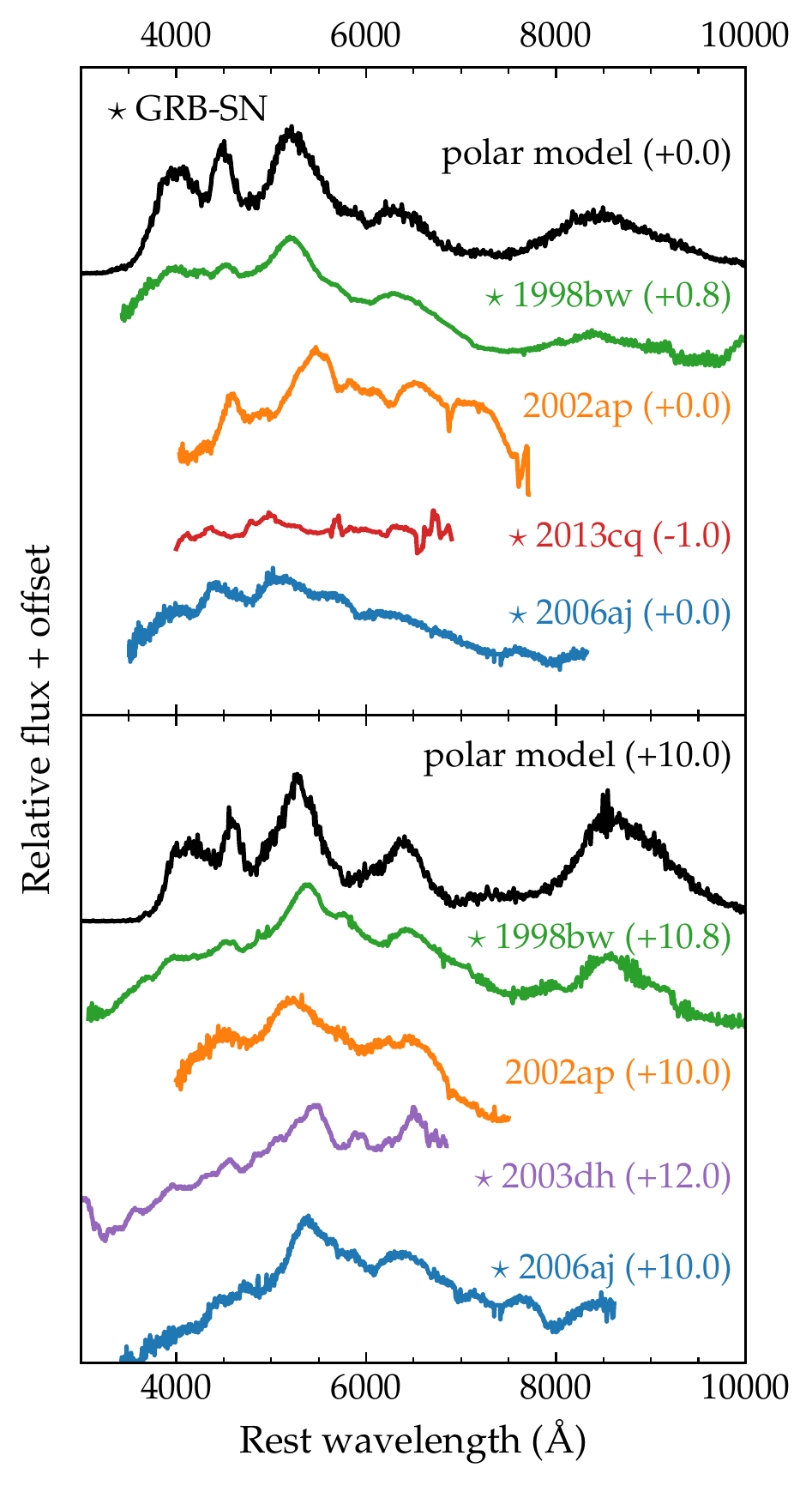}
\caption{The spectrum of our model (seen pole-on) at peak and ten days after peak, compared to observed SNe Ic-BL at similar phases. SNe with a confirmed corresponding GRB are marked with a star. Our model has line widths comparable to SNe classified as Ic-BL, though it is bluer than some of the observed SNe, and has more prominent features in the UV.}
\label{fig:spec.pk10}
\end{figure}

\begin{figure}\includegraphics[width=3.5in]{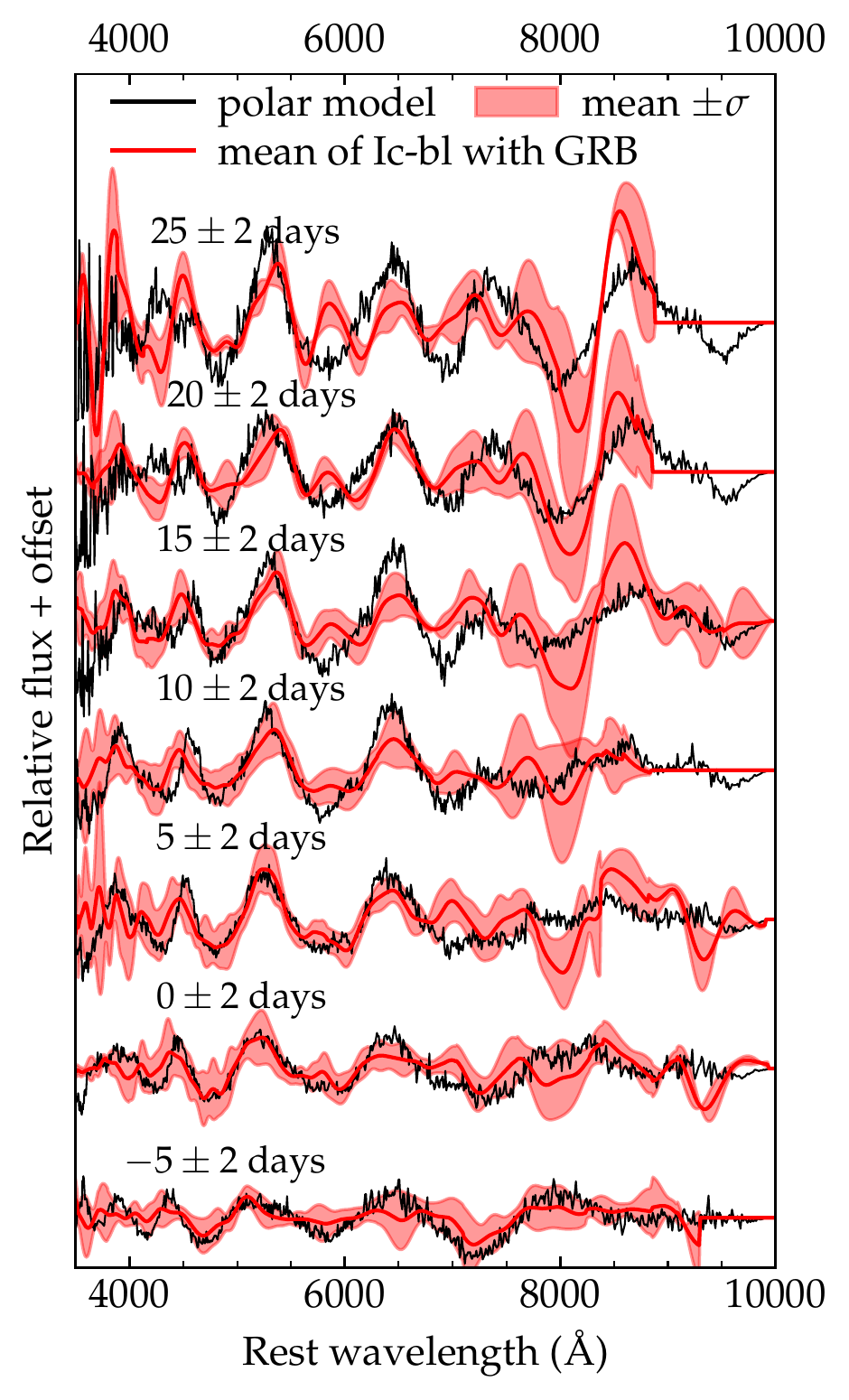}
\caption{The polar model spectrum at multiple phases compared to average spectra of SNe Ic-BL with a coincident GRB, calculated by \citet{Modjaz_etal_2016_GrbSne}. All spectra are continuum-subtracted using the methods of \citet{Modjaz_etal_2016_GrbSne} and \citet{Liu.Modjaz.ea_2016_se.sn.spectra}. The averages are plotted as red curves, and the light red bands show the region within one standard deviation of the mean. Our model generated spectra with stronger features than the average spectra, so we scaled average spectra and the standard deviations by a factor of 3 to allow an easier comparison of the widths of the absorption features in the two cases.}
\label{fig:avspec}
\end{figure}

Figures \ref{fig:spec.pk10} and \ref{fig:avspec} compare our model's polar spectrum to data, and further demonstrate the commonalities between the model and observed SNe Ic-BL.
In Figure \ref{fig:spec.pk10}, we present the spectrum of the model at peak and ten days after peak, alongside spectra of observed SNe Ic-BL, both with and without coincident GRBs, at comparable phases.
Figure \ref{fig:avspec} shows the time evolution of the polar model spectrum, with the continuum removed, relative to the average flattened spectra for SNe Ic-BL with associated GRBs calculated by \citet{Modjaz_etal_2016_GrbSne}. The continuum subtraction of the model was carried out using the same data-driven procedure applied to observed spectra 
(see \citealt{Modjaz_etal_2016_GrbSne} and \citealt{Liu.Modjaz.ea_2016_se.sn.spectra} for details.)
To concentrate on the comparison of the spectral line features of our model with those of the statistical data set of SNe Ic-BL with associated GRBs, we have scaled the average spectra.
We also analyze the synthetic polar spectrum using SNID.
References for spectra in Figure \ref{fig:spec.pk10} are given in Table~\ref{tab:specsrc}.

\begin{table}[h]
\caption{Summary of Comparison Spectra}\label{tab:specsrc}
\begin{threeparttable}
\centering
\begin{tabularx}{\columnwidth}{*{3}Y}
\textit{SN} & \textit{Classification} & \textit{References}\tnote{a} \T\B \\
\hline
1998bw & Ic-BL, with GRB & G98, P01 \T\T \\
2002ap & Ic-BL, w/o GRB & SNID\tnote{b} (G02, F03), M14 \\
2003dh & Ic-BL, with GRB & S03, M03, K03, K04, D05 \\
2006aj & Ic-BL, with GRB & M06, P06, M14 \\
2013cq & Ic-BL, with GRB & X13 \\
\hline
\end{tabularx}
\begin{tablenotes}
\item[a] References: G98--- \citet{Galama98_bwDisc}; P01 --- \citet{Patat.ea_2001_98bw.data}; G02 --- \citet{GalYam.ea._2002_02ap.data}; F03 --- \citet{Foley.ea_2003_2002ap.data}; M14 --- \citet{Modjaz.ea_2014_se-sn.dat}; S03 --- \citet{Stanek.ea_20032003dh.data}; M03 --- \citet{Matheson.ea_2003_2003dh.data}; K03 --- \citet{Kawabata.ea_2003_2003dh.data}; K04 --- \citet{Kosugi.ea_2004_2003dh.neb.data}; D05 --- \citet{Deng.ea_2005_2003dh.data}; M06 --- \citet{Modjaz.ea_2006_2006aj.data}; P06 --- \citet{Pian.ea_2006_2006aj.data}; X13 --- \citet{Xu.ea_2013_2013cq.dat};
\item[b] SNID --- in SNID release version 5.0 with templates-2.0 by \citet{Blondin.Tonry_2007_SNID.paper}, with original references in parenthesis
\end{tablenotes}
\end{threeparttable}
\end{table}

The model successfully reproduces the major characteristics of Ic-BL spectra, which result from the blending of highly Doppler-broadened line profiles. 
The similarity of the model spectra to observations and to mean spectra can be seen in
Figures \ref{fig:spec.pk10} and \ref{fig:avspec}. 
SNID also classified the model spectra as belonging to the Ic-BL class, though the individual spectra SNID matched to the model at different phases, summarized in Table~\ref{tab:snid}, tended to be GRB-absent SNe Ic-BL, which \citet{Modjaz_etal_2016_GrbSne} showed to have systematically narrower lines than GRB-SNe. 
However, Figure \ref{fig:avspec} shows that the line widths of the model are generally comparable to the mean spectra of SNe Ic-BL with associated GRBs, so the interpretation of the SNID matches is not straightforward. What is clear is that the model belongs unambiguously to the Ic-BL category based on its broad lines and strong line-blending.

\begin{table}[h]
\caption{Best-match spectra from SNID}\label{tab:snid}
\begin{threeparttable}
\centering
\begin{tabularx}{\columnwidth}{Y|*{3}Y}
\multicolumn{1}{c}{\textbf{Polar model}} & \multicolumn{3}{c}{\textbf{SNID match}} \T\B \\
\textit{Phase} & \textit{SN} & \textit{Phase} & \textit{Classification} \\ 
\hline \\
-5 days & 2007ru & $-$3 days & Ic-BL, w/o GRB \T \\
+0 days & PTFgzk & $-$1 day & Ic-BL w/o GRB \\
+0 days & 2012bz & +1 days & Ic-BL with GRB \\
+10 days & 2007I & $>10$ days\tnote{a} & Ic-BL w/o GRB \\
\hline
\end{tabularx}
\begin{tablenotes}
\item[a] SN 2007I has no confirmed light-curve peak, so the phase of that supernova's spectrum is not precisely known.
\end{tablenotes}
\end{threeparttable}
\end{table}

As shown in Figures \ref{fig:spec.pk10}, the model spectra are bluer than most of the observed SNe.
The strengths of individual spectral features are also not always well captured. In particular, the model exhibits prominent spikes in the blue and UV that are mostly absent from observed SNe. Narrow Fe features at these wavelengths were also present in the spectra of \cite{Dessart2017_asymBLics}, who carried out a systematic survey of 1D Ic-BL models using the non-LTE transport code \texttt{CMFGEN} \citep{Hillier.Dessart_2012_cmfgen.code}. In particular, their model r6e4BH, which has an ejecta mass and energy close to our model, appears at phase 8.3 days to be similarly spiky for $\lambda \lesssim 5000$ \AA\ (their Figure 3). 

The over-pronounced blue and UV features are also apparent when comparing to average Ic-BL spectra. While the model and the mean spectra have absorption features of similar width, the model's features are much stronger than the means'; the averages (red lines) and standard deviations (light red bands) in Figure \ref{fig:avspec} were scaled by a factor of 3. This suggests that the energetics of the model are roughly correct, but the composition, which we have not rigorously investigated here, may need to be adjusted to improve agreement with observation. For example, incorporating a more realistic spatially-dependent composition could improve the situation.

\section{Discussion and Conclusion} \label{sec:disc}

We have used sophisticated numerical SRHD calculations, radiation transport, and data analysis techniques to investigate the possibility that a GRB engine, without any additional energy source, can produce both a narrow, highly relativistic jet and a more isotropic supernova explosion with the spectral characteristics typical of GRB-SNe. This study provides an end-to-end description of a jet-driven core-collapse event.

We have shown that a GRB engine injected into a stripped-envelope star not only produces a long-duration GRB, it can also transfer enough energy to the stellar material to unbind it, resulting in a SN explosion.
The engine also heats a portion of the ejected material to high enough temperatures that the synthesis of \nickel\ is favored by nuclear statistical equilibrium. 

Our fiducial GRB engine and progenitor models trigger a SN typical of SNe Ic-BL. The SN's bolometric light curves have luminosities and shapes consistent with observed SNe Ic-BL. The model spectra have the broad absorption features that underpin the Ic-BL classification. We considered only a single engine-progenitor pair for this study. However, a broader parameter survey is planned for the future, and it may identify systems that better match average SNe Ic-BL properties, or that mimic individual unusual Ic-BL events (e.g. 1998bw). 

We have also explored the effect of asymmetry on the SN observables.
Our model creates a mildly asymmetric ejecta 
that imparts a slight viewing-angle dependence to the light curves, and a moderate dependence to the spectra. At very early times, spectral features observed along the pole are broadened and blue-shifted relative to spectra observed from an equatorial vantage.
However, this difference fades with time, and it is not clear whether an observer located on-axis would infer a meaningfully higher photospheric velocity or kinetic energy than an observer near the equator.

We have established that a single engine can produce the observed properties of both a long GRB and a highly kinetic SN. 
However, much remains to be done. Future work (in prep.) will explore in more detail the effects of progenitor models and engine parameters on both the relativistic jet and the supernova produced, and delineate how the the production of \nickel, the kinetic energy of the explosion, the degree of asphericity in the ejecta, and the breadth of the lines depend on these parameters. 
We will also extend our calculations to late times to see whether the aspherical ejecta geometries produced in jet-driven explosions can self-consistently explain both the early- and late-time spectra and light curves of SNe Ic-BL.

\acknowledgments

We are grateful to Io Kleiser and Eliot Quataert for helpful comments and discussions.

Y.-Q. Liu is supported in part by NYU GSAS Dean's
Dissertation Fellowship.  M. Modjaz and the SNYU
group are supported by the NSF CAREER award AST-
1352405 and by the NSF award AST-1413260.

This project is supported in part by a Department of Energy Office of Nuclear
Physics Early Career Award, and by the Director, Office of Energy
Research, Office of High Energy and Nuclear Physics, Divisions of
Nuclear Physics, of the U.S. Department of Energy under Contract No.
DE-AC02-05CH11231, and from NSF grant AST-1206097

This work made use of the data products generated by the NYU SN group, and 
released under DOI:10.5281/zenodo.58766, 
available at \url{https://github.com/nyusngroup/SESNtemple/}.

\bibliographystyle{apj} 
\bibliography{refs}

\end{document}